\documentclass[twocolumn,prb,10pt,aps,longbibliography,superscriptaddress]{revtex4-2}
\usepackage{graphicx}
\usepackage{float}
\usepackage{amsmath}
\usepackage{amssymb}
\usepackage{amsfonts}
\usepackage{dsfont}
\usepackage{dcolumn}
\usepackage{bbm}
\usepackage{bm}
\usepackage{tikz}
\usetikzlibrary{positioning}
\usepackage{mathtools}
\usepackage{braket}
\usepackage{physics}
\usepackage{siunitx}
\usepackage[unicode]{hyperref}
\usepackage[nameinlink]{cleveref}

\usepackage{bbold} 
\usepackage{nicefrac}
\usepackage{tikz}
\usetikzlibrary{arrows.meta}
\usepackage{xcolor}

\usepackage[makeroom]{cancel}
\crefname{figure}{Fig.}{Figs.}

\newcommand{\bes}{\begin{subequations}}
\newcommand{\ees}{\end{subequations}}
\newcommand{\be}{\begin{equation}}
\newcommand{\ee}{\end{equation}}

\newcommand{\sumnn}{\sum_{\langle i,j\rangle}}
\newcommand{\kk}{\mathbf{k}}
\newcommand{\lr}[1]{\left( #1 \right)}

\newcommand{\lrg}[1]{\left\{#1\right\}}
\newcommand{\lara}[1]{\langle #1 \rangle}

\newcommand{\nbi}[1]{#1_i^\dagger {#1}_i}
\newcommand{\nbk}[1]{#1_\mathbf{k}^\dagger {#1}_\mathbf{k}}

\newcommand{\sumk}{\sum_{\mathbf{k}}}

\newcommand{\ak}{a_\mathbf{k}}
\newcommand{\bk}{b_\mathbf{k}}
\newcommand{\akm}{a_\mathbf{-k}}
\newcommand{\bkm}{b_\mathbf{-k}}

\newcommand{\gam}{\gamma_\mathbf{k}}

\begin{document}
    \title{Switching magnetization of quantum antiferromagnets: \\
    Schwinger boson mean-field theory compared to exact diagonalization}


    \author{Florian Johannesmann}
    \email{florian.johannesmann@tu-dortmund.de}
    \affiliation{Condensed Matter Theory, 
    TU Dortmund University, Otto-Hahn-Stra\ss{}e 4, 44227 Dortmund, Germany}

    \author{Asliddin Khudoyberdiev}
    \email{asliddin.khudoyberdiev@tu-dortmund.de}
    \affiliation{Condensed Matter Theory, 
    TU Dortmund University, Otto-Hahn-Stra\ss{}e 4, 44227 Dortmund, Germany}

    \author{G\"otz S.\ Uhrig}
    \email{goetz.uhrig@tu-dortmund.de}
    \affiliation{Condensed Matter Theory, 
    TU Dortmund University, Otto-Hahn-Stra\ss{}e 4, 44227 Dortmund, Germany}

    \date{\textrm{\today}}

\begin{abstract}
Antiferromagnets have attracted significant attention because of their considerable potential in engineering high-density and ultrafast memory devices, a crucial and increasingly demanded component of contemporary high-performance information technology. Theoretical and experimental investigations are actively progressing to provide the capability of efficient switching and precise control of the Néel vector, which is crucial for the intended practical applications of antiferromagnets. Recently, a time-dependent Schwinger boson mean-field theory has been successfully developed to study the sublattice magnetization switching in anisotropic quantum antiferromagnets [K. Bolsmann $et \, al.$, \textcolor{blue}{\hyperlink{10.1103/PRXQuantum.4.030332}{PRX Quantum $\mathbf{4}$, 030332 (2023)}}]. Here we use a complementary  exact diagonalization method to study such sublattice magnetization switching, but in small-cluster quantum antiferromagnets, by means of an external magnetic field. Furthermore, this article aims to support the findings of the  Schwinger boson approach. We show that the results of both approaches are consistent at short time scales, with only about 12.5 $\%$ deviations. The consistency of the outcomes obtained through this alternative exact approach demonstrates that the time-dependent Schwinger boson mean-field theory is a versatile framework to capture the essentials of the switching process in quantum antiferromagnets. Thereby, the findings of current article  pave the way for further theoretical and computational progress in the study of antiferromagnets for engineering spintronic devices with ultrahigh density and ultrafast speed.
\end{abstract}

    \maketitle

\section{Introduction}	

The manipulation and control of magnetization in magnetic materials is of great interest in many areas
of physics due to their applications for high-density memories and spintronic devices \cite{jungw16,baltz18}. Moreover, experimental evidence suggests that antiferromagnetic materials combined with heavy metal layers can be used as spintronic terahertz (THz) emitters \cite{qiu21,rongio23}. Therefore, antiferromagnets have been proposed as a
promising platform because of their ultrafast spin dynamics compatible with ultrafast information processing, robustness against external noise fields and absence of stray fields between the domains \cite{macdon11,gomon17,jungf18}.  Research is currently underway to determine the viability of employing antiferromagnets for information storage and processing \cite{jungw18,kimel24}. 

In antiferromagnets, the absence of net magnetization poses a substantial obstacle to the effective readout and manipulation of the direction of the N\'eel vector. However, intense theoretical \cite{gomon14,gomon16,bolsm23,khudoy24,khudoy25,yarmo25}  and experimental \cite{zelez14,zelez17,bodna18,behov23} research is currently underway, with significant progress in recent years to demonstrate the possibility of efficient switching \cite{wadle16,behov23,han2024,guo2025}, read-out \cite{bommana21,grigo21,kimel24} and full control \cite{ghara25, polew25,toyoda25} of the antiferromagnetic order, to make it practically applicable.

The experimental observation of deterministic electrical 180$^\circ$ switching of the N\'eel vector in the spin-split antiferromagnet Mn$_5$Si$_3$ was achieved through utilization of a spin-orbit torque with an asymmetric energy barrier \cite{han2024}. In this experiment, the read-out of the N\'eel vector switching is realized through the flipping the sign of the anomalous Hall conductivity.
This even made the fabrication of an antiferromagnetic device possible
with both high ("0" state) and low ("1" state) resistance, thereby enabling the realization of robust write and read-out cycles  \cite{han2024}. Moreover, atomistic spin simulations confirm 
the considered switching processes \cite{han2024}. Similarly,  current-induced spin-orbit torques enabled full switching of the N\'eel vector in  antiferromagnet (Fe,Co)$_3$GaTe$_2$ with perpendicular magnetic anisotropy \cite{guo2025}. Very recently, Jourdan $et\, al.$ \cite{jourd25} even  highlighted experimentally two distinct mechanisms using a current pulse for switching the sublattice magnetization in the antiferromagnet Mn$_2$Au. The authors concluded that current-induced heating (thermomagnetoelastic effect) drives the N\'eel vector switching under 10 $\mu s$ pulses, while the N\'eel spin-orbit-torque \cite{bodna18,chen2019,behov23} is responsible for complete reorientation under pulses shorter than 100 ns, as required for ultrafast applications. Consequently, one can conclude that the state-of-the-art antiferromagnetic switching has undergone significant progress in the experimental domain.

Clearly, this calls for further theoretical support. So far, the 
switching mechanism was primarily studied for classical vectors
by the Landau-Lifshitz-Gilbert equation \cite{gomon10,gomon16} 
which can be linked to the quantum Lindblad formalism \cite{uhrig25}
or similarly by atomistic spin dynamics simulations \cite{ross24,han2024}
which include thermal fluctuations on a stochastic level.
Our objective is to raise the theoretical description to 
the quantum level. In equilibrium, spin wave theory at face value
or in a self-consistent way allow one to incorporate the leading
quantum corrections. The standard choices of the Holstein-Primakoff
representations \cite{holst40} or Dyson-Maleev representation \cite{dyson56a,malee58} 
only capture the fluctuations around a static, prechosen order.
Thus, they cannot capture strong deviations such as they
occur in the re-orientation of the magnetization, except 
in very tedious considerations \cite{ruckr12}.
Thus, we choose the Schwinger boson representation \cite{schwi52}.

\begin{figure}
 \begin{tikzpicture}[
    every node/.style={
        font=\fontsize{11}{11}\selectfont, 
    }
]
\node at (-2.4,1.3) {initial state};
\node at (3.0,0.2) {final state};
\node at (-3.6,0.4) {x};
\node at (-3.4,-0.2) {z};
\node at (-4.05,0.9) {y};
\node at (-0.3,1.55) {$\vec{h}$};
\node at (0.95,1.36) {$\vec{h}$};
\node at (0,0) {\includegraphics[width=\columnwidth]{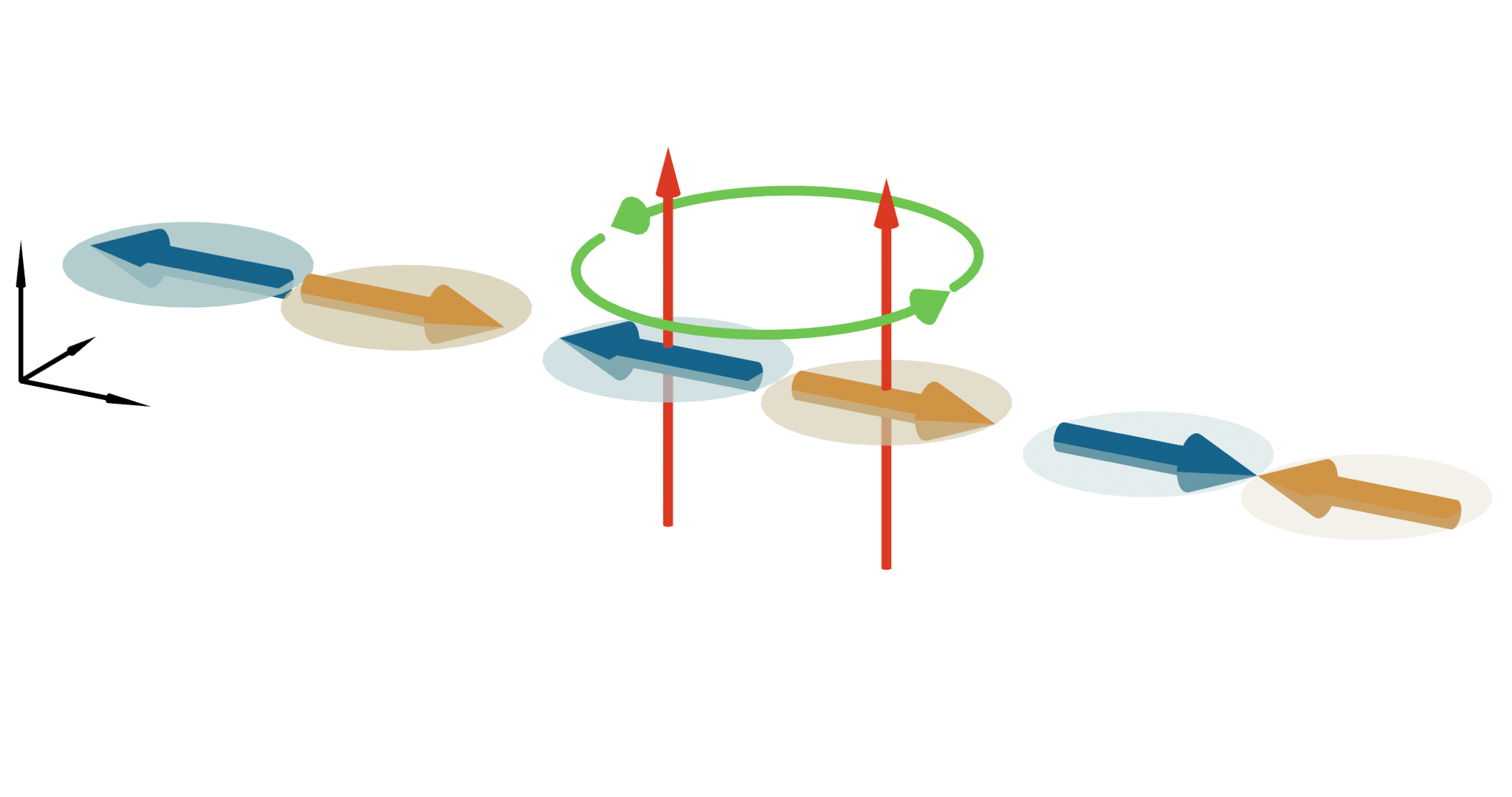}};
\end{tikzpicture}
\vspace{-1.7cm}
\caption{Illustration of the manipulation of the sublattice magnetization. The initial state is shown on the left with two linked sublattices. The arrows in the circles show the sublattice magnetization direction. A uniform magnetic field is applied to switch the order (shown in the middle) and the final switched state is obtained (shown on the right). Round green arrows indicate the Larmor precession of the sublattice spins.} \label{fig:switch}
\end{figure}
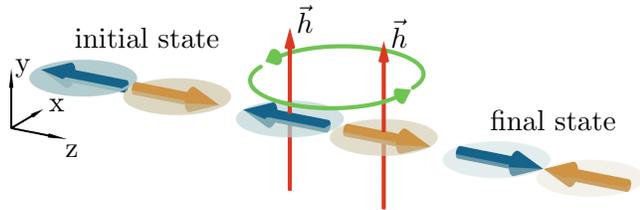

 The time-dependent Schwinger boson mean-field theory (SBMFT) has been developed to study switching in quantum antiferromagnets driven by external magnetic fields \cite{khudoy25,bolsm23,khudoy24,khudoy25scipost}. This approach successfully demonstrated that control of the N\'eel vector can be achieved by applying strong uniform fields (\cref{fig:switch}) \cite{bolsm23,khudoy24}. It was shown that the size of the spin gap in the dispersion determines the strength of the uniform external field required for the switching. Time-dependent control fields allow for the reduction of the strength of the required switching field due to resonant control \cite{khudoy24}. Moreover, staggered magnetic fields on neighboring sublattices generate exchange field enhancement, so that switching occurs for significantly lower fields \cite{khudoy25}, similar to the effect of N\'eel spin-orbit torque.  
 
 Even though the quantum system is treated to be a closed system, the dynamics of sublattice magnetization after
 switching is not fully coherent, but a slow decay of the oscillations is observed. This phenomenon is due to  dephasing caused by the numerous modes
 at different frequency in the system \cite{khudoy25,bolsm23,khudoy24,khudoy25scipost}. Furthermore, considering an open quantum system by including spin-lattice relaxation, derived from the Lindblad formalism \cite{breue06}, an exponential decay of the oscillations and fast convergence to the steady-state after full ultrafast switching have been observed \cite{khudoy25scipost}. Therefore, the methodological progress to deal with Schwinger mean-fields including Lindblad dissipators facilitated also to address the differences between dephasing and spin-lattice relaxation in true switching processes. But still, to validate the results derived so far either classically or by
 quantum mean-field approaches, the use of additional quantum approaches is necessary.
 
In this work, we report results obtained from the complementary exact diagonalization (ED) method in small clusters to simulate switching processes in antiferromagnets by means of external magnetic fields. We consider different two-dimensional antiferromagnetic clusters with spin-1/2 sites, as approximations for the infinite two-dimensional square lattice, see below. 

We emphasize that the comparison of mean-field results to
 small cluster ED results is an intricate task:
 The mean-field approach easily allows to describe rather
 large systems and there is little difficulty to capture
 symmetry breaking. The ED approach, however, is limited to
 rather small systems of $\mathcal{O}(20)$ spins which do not
 show any symmetry breaking due to quantum tunneling, see for
 instance Ref. \cite{beckm17}.
 To generate sublattice magnetization which makes it possible
 to address the switching process at all, some initial state preparation 
 with small staggered fields cannot be circumvented.

The paper is organized in the following manner. In section \ref{sec:methods}, we provide a brief description of time-dependent SBMFT and a detailed discussion of the exact dioganalization method. Section \ref{sec:results} reports the results of the exact diagonalization method and its comparisons with the findings of SBMFT. The conclusions and
perspectives of our work are reported in section \ref{sec:concl}. The details about the initialization of systems and  the structure of clusters are provided in the Appendix.

\section{Model and Method} \label{sec:methods}

\begin{figure}
    \centering
    \includegraphics[width=\columnwidth]{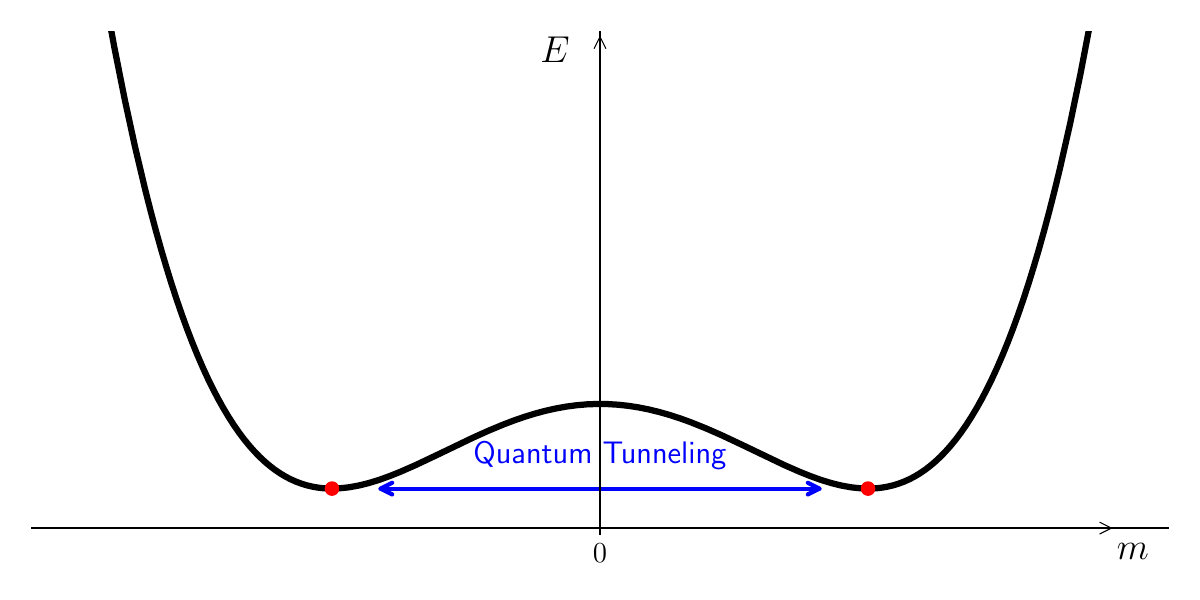}
    \caption{Illustration of quantum tunneling between two ordered states, depicted as red dots, 
    separated by an intermediate activation energy, drawn as a solid black line.}
    \label{fig:quantum-tunneling}
\end{figure}

We start with a system described by the  Heisenberg model on a square lattice with Hamiltonian 
\begin{equation}\label{eq:anizHam}
\mathcal{H}_\text{0}=J \sum_{\langle i,j\rangle}\left[\frac{\chi}{2}(S_i^+S_j^-+S_i^-S_j^+)+S_i^zS_j^z\right]- h_z \sum_{i} (-1)^{i} S_{i}^{z} ,
\end{equation}
where $\chi$  is an anisotropy parameter, $J$ the exchange coupling, and $h_z=g\mu_BB_z$ an external field in energy units. 
Here, $g$  is the Land\'e $g$-factor, $\mu_B$  is the Bohr magneton, and $B_z$  is the external magnetic field along the  $z$ axis. Subsequent analyses of the results employ energies measured in units of the antiferromagnetic exchange coupling $J>0$.  We want to perform exact diagonalization calculations on small clusters to corroborate the SBMFT results on switching \cite{bolsm23} by capturing the main quantum effects at zero temperature. Therefore, we included a Zeeman term with an staggered tiny auxiliary field  in the $z$ direction in Eq.\ \eqref{eq:anizHam} to force  antiferromagnetically alignment of the spins \cite{lusch09}.  
Otherwise, the total spin of the ground state is zero and no spin direction is singled out
\cite{lieb62}.
Thus, the auxiliary field is necessary for small cluster sizes to obtain a ground state with proper finite values of sublattice magnetization \cite{caci24} by counteracting  quantum tunneling
between the two states of long-range order in easy-axis spin systems, see Fig.\ \ref{fig:quantum-tunneling} and for instance Ref.\ \cite{beckm17}.
Even for a robust  convergence of self-consistency equations of the mean-field the auxiliary field is indicated \cite{auerb94} to determine the orientation of the symmetry breaking \cite{bolsm23}. 
For diverging system size, the necessary auxiliary fields vanish.

Below, we will briefly recap SBMFT, then provide details of our ED approach.

\subsection{Schwinger boson mean-field approximation} 
\label{subsec:sbmft}

The switching of the antiferromagnetic order implies more than small fluctuations around the ordered state. In order to analyze the switching processes in antiferromagnets, the SU(2) Schwinger boson representation is the most suitable \cite{bolsm23,khudoy24,khudoy25,khudoy25scipost} due to its capability to describe spin-symmetric phases as well \cite{auerb94}.  In the SU(2) Schwinger boson representation, the spin operators of the Heisenberg Hamiltonian are replaced by two flavors of bosonic operators as 
\be
\mathbf{S}_i=\frac{1}{2}\begin{pmatrix}
a^\dagger_i  b^\dagger_i
\end{pmatrix}\boldsymbol{\sigma}
\begin{pmatrix}
  a_i \\
  b_i \\
\end{pmatrix}, 
\ee
where $\boldsymbol{\sigma}$ is the Pauli vector. To perform a meaningful mapping of the physical spin Hilbert space to the subspace of the bosonic Hilbert space, one limits the Schwinger boson occupation at each site to be

\be \label{eq:constr}
\nbi{a}+\nbi{b}=2S,
\ee
where $S$ is the spin length. Then, two Schwinger bosons span the physical Hilbert space for $S=1/2$ as $\ket{1,0}:=\ket{\uparrow}$ and $\ket{0,1}:=\ket{\downarrow}$. Other linear combinations of the bosons can represent other spin orientations in the sublattice. The sublattice magnetization is calculated by
\begin{equation}
\label{eq:magnetization}
   m_z =\langle S_i^z\rangle= \frac{1}{2}\lr{\langle \nbi{a} \rangle  - \langle \nbi{b}\rangle}. 
\end{equation}
Prior to the substitution of the spin operators with bosonic operators in Eq. \eqref{eq:anizHam}, it is advantageous to introduce the bond operators as $A_{ij}:=a_ib_j-b_ia_j$ and $B_{ij}:=a_ib_j+b_ia_j$, which enable rewriting the Hamiltonian in a compact form as 

\begin{align}
  \mathcal{H}_0 &= -\frac{1}{4}\sumnn\lrg{(1+\chi)A_{ij}^\dagger A_{ij}+(1-\chi)B_{ij}^\dagger B_{ij}-4S^2 } \nonumber
  \\
  &-\frac{h_z}{2}\sum_i(-1)^i(\nbi{a}-\nbi{b}).
	\label{eqn:Hbond}
\end{align}
The bond operator $A_{ij}$ is an antisymmetric to the interchange of $i\rightarrow j$. To simplify analytical calculations, we perform a 180$^\circ$ spin rotation on one sublattice about the $y$ axis which amounts to a unitary transformation  not affecting physics. As a result, the bond operators become symmetric as $A_{ij}:=a_ia_j+b_ib_j$ and $B_{ij}:=a_ia_j-b_ib_j$, and we retain  the full translational  invariance of the  Hamiltonian. The alternating field $h_z$ also becomes a uniform field, which simplifies the calculations significantly. Additionally, the mean-field approximation - the primary tool employed in the subsequent subsection, becomes compactly applicable even in the SU(2) representation, e.g., $A_{ij}^\dagger A_{ij}\approx A_{ij}^\dagger \langle A_{ij}\rangle +A_{ij} \langle A_{ij}^\dagger\rangle-\langle A_{ij}^\dagger\rangle\langle A_{ij}\rangle$ \cite{auerb94,rayki93, pires21}. This scheme is justified by an expansion in the inverse number of flavors, here two. It is not identical to Wick's theorem, though bearing similarities.  

The mean-field Hamiltonian in momentum space reads 

\begin{align}
   &\mathcal{H}_0^\text{MF} = E_0 -\frac{1}{2}\sumk \gam \big(C_-\ak^\dagger \akm^\dagger 
	+ C_+\bk^\dagger\bkm^\dagger +C_-^*\ak\akm  \nonumber
	\\
	& +C_+^*\bk\bkm\big) + \Big(\lambda-\frac{h_z}{2}\Big) \sumk \nbk{a} + \Big(\lambda+\frac{h_z}{2}\Big)\sumk\nbk{b},
	\label{MFHam}
\end{align} 
where $C_{\pm}:= A(1+\chi)\mp B(1-\chi)$ and $C_{\pm}^*:= A^*(1+\chi)\mp B^*(1-\chi)$ with $A:= \lara{a_i a_j + b_i b_j}$,  $B:= \lara{a_i a_j - b_i b_j}$ and  $A^*:= \lara{a_i^\dagger a_j^\dagger + b_i^\dagger b_j^\dagger}$,  $B^*:= \lara{a_i^\dagger a_j^\dagger - b_i^\dagger b_j^\dagger}$ . In our notation the parameters $C_-$ and $C_+$ correspond to the bosons $a$ and $b$, respectively. The constant $E_0$ is a constant energy and $\gam=\frac{1}{4}\sum_\delta e^{i\kk \cdot\boldsymbol{\delta}}$. The Lagrange term with the Lagrange parameter $\lambda$ is included in the Hamiltonian to fix the average number of bosons per lattice site according to the constraint in 
Eq.\ \eqref{eq:constr}. 

Next, we  diagonalize the mean-field Hamiltonian using Bogoliubov transformations and obtain the dispersion relations for the bosonic flavors

\begin{equation} \label{eq:disper}
\omega_\mathbf{k}^\pm=\sqrt{\Big(\lambda\pm\frac{h_z}{2}\Big)^2-(\abs{C_\pm}\gam)^2}.
\end{equation}
Because of simulations of finite size clusters, both bosons acquire an energy gap as
\begin{equation} \label{eq:engap}
    \Delta^\pm=\omega_{\mathbf{k}=0}^\pm.
\end{equation}
The closed set of self-consistent mean-field equations 
is constructed using the mean-field parameters and the constraint in Eq.\ \eqref{eq:constr} as
\bes \label{eq:selfconsistent}
     \begin{align}
    A&=\lara{a_ia_j}+\lara{b_ib_j}=\frac{1}{N}\sumk\gam\left(\lara{\ak \akm}+\lara{\bk \bkm}\right),
    \\
    B&=\lara{a_ia_j}-\lara{b_ib_j}=\frac{1}{N}\sumk\gam\left(\lara{\ak \akm}-\lara{\bk \bkm}\right),
    \\
    2S&=\lara{\nbi{a}}+\lara{\nbi{b}}=\frac{1}{N}\sumk\left(\lara{\nbk{a}}+\lara{\nbk{b}}\right),
    \end{align}
\ees
where 
\bes \label{eq:expectvalues}
   \begin{align}
        \lara{\nbk{a}}&=\frac{1}{2}\left(\frac{\lambda-\frac{h_z}{2}}{\omega_\kk^-}-1\right),
    \\
    \lara{\nbk{b}}&=\frac{1}{2} \left(\frac{\lambda+\frac{h_z}{2}}{\omega_\kk^+}-1\right),
    \\
    \lara{\ak \akm}&=\frac{C_-\gam}{2\omega_\kk^-},
    \\
    \lara{\bk \bkm}&=\frac{C_+\gam}{2\omega_\kk^+}.
    \end{align}
\ees
We prepare the initial state such that there are more $a$ bosons than $b$ 
bosons \cite{bolsm23}, i.e, the initial sublattice magnetization is finite and positive. The self-consistent equations \eqref{eq:selfconsistent} 
are solved to find the mean-field parameters and the Lagrange parameter, which completes the initialization. 

The system is driven through the magnetic field along the $y$ axis to reach the reorientation of the N\'eel vector 
\be
\mathcal{H}_\mathrm{m}=-h_y\sum_iS^y_i=- \frac{h_y}{2i} \sumk\big(\ak^\dagger\bk + \bk^\dagger\ak\big),
\ee
where $h_y=g\mu_BB_y$. The above Hamiltonian does not commute with the mean-field Hamiltonian in 
Eq.\ \eqref{MFHam}. Therefore, the time evolution of the expectation values of bilinear operators are computed using the Heisenberg equation of motion, and a closed set of differential equations are constructed using the full mean-field Hamiltonian $\mathcal{H}_\mathrm{t}=\mathcal{H}_0^\mathrm{MF}+\mathcal{H}_\mathrm{m}$ 
\cite{bolsm23, khudoy24, khudoy25}. Note that also the Lagrange parameter $\lambda$
needs to made time-dependent to keep the constraint fulfilled in the course of the 
evolution \cite{fause24,khudoy25}.
Lastly, the differential equations are solved at each  momentum point $\mathbf{k}$ in the Brillouin zone, where the solutions enable the analyzes of the dynamics of the magnetization as
\begin{align} \label{eq:magdynam} \nonumber
    m_z(t)&=\frac{1}{2}\lr{\langle \nbi{a} \rangle  - \langle \nbi{b}\rangle}\\
    &=\frac{1}{2N}\sumk\left(\lara{\nbk{a}}-\lara{\nbk{b}}\right),
\end{align}
with $N$ denoting the number of lattice sites.

\subsection{Exact diagonalization}

For the exact diagonalization (ED) we start from the Hamiltonian in Eq. \eqref{eq:anizHam}.
In order to take full advantage of the translational symmetry of the underlying lattice, one sublattice, designated as $j$, is rotated about the spin $S_y$ axis as for the mean-field treatment. The rotated Hamiltonian now reads
\begin{equation}
\label{eq:anizHam-with-hz-field-rotated}
\mathcal{H}_R = -J \sum_{\substack{<i,j>}} \left[ \frac{\chi}{2} \left(S_i^+ S_j^+ + S_i^- S_j^- \right) + S_i^z S_j^z \right] - h_z \sum_{i} S_i^z .
\end{equation} 
Regarding the system size, we use translational symmetries in order to reduce the effective Hilbert space dimension to reach a maximum number of $24$ spins. For the simulations, we use the software package QuSpin \cite{weinberg17, weinberg19} that supports symmetry operations and efficient run time and memory performance. The ground state search is done by using a sparse eigen-solver that runs an implicitly restarted Lanczos algorithm within the subspace $H_{(k_x, k_y)}$ with momentum quantum numbers $(k_x = 0, k_y = 0)$. For the time evolution of the magnetization, we solve the time-dependent Schrödinger equation by using a high order Runge Kutta method. We compare the short-time magnetization dynamics and the required switching fields at different easy-axis anisotropy strengths to the results of the SBMFT. The correspondence of the results of two different approaches allows for a cautious confirmation of the results of time-dependent SBMFT \cite{bolsm23,khudoy24,khudoy25}. 

\begin{figure}
    \centering
    \includegraphics[width=\columnwidth]{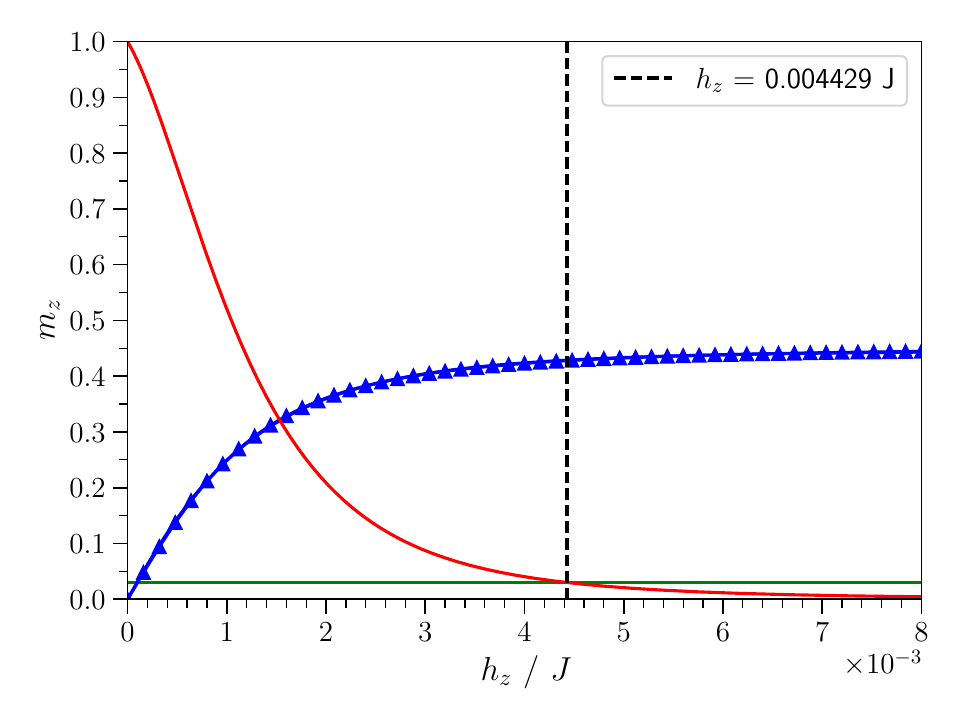}
    \caption{Sublattice magnetization values (blue dots) dependent on the $h_z$ field for a $4\times4$ cluster at $\chi = 0.6$, 
    fitted with the hyperbolic ansatz \eqref{eq:FitFunction} (solid blue curve). 
    The normalized derivative is shown as a red solid line and the threshold value of $0.03$ is depicted as a green horizontal line. The estimated initial $h_z$ field is shown as a vertical dotted black line.}
    \label{fig:magnetization-dependent-on-hz}
\end{figure}

The first step  is to determine a suitable ground state in the method that can serve as an initial state with finite sublattice magnetization for the time evolution. Therefore, a suitable $h_z$ field is necessary and must be found in the first place. According to Marshall's theorem \cite{auerb94}, the ground state of the Hamiltonian is located in the symmetry reduced subspace $H_{(k_x, k_y)}$ with momentum quantum numbers $(k_x = 0, k_y = 0)$. For several $h_z$ values, the ground state and the corresponding magnetization of one spin of the lattice is calculated. The obtained values are then fitted with a hyperbolic ansatz for the magnetization curve
\begin{equation}
\label{eq:FitFunction}
m_z(h_z) = a \cdot \tanh{\left( \frac{b \cdot h_z}{1 + c \cdot h_z} \right)} ,
\end{equation}
where $a, b$ and $c$ are fit parameters. The parameter $a$ can be thought of as the magnetization reached at an infinitely large field. A suitable value for $h_z$ is obtained by setting the  normalized derivative of the fit function to 0.03 as
\begin{equation}
    m_z^{\prime}(h_z) = \frac{1}{\cosh{\left(\frac{b \cdot h_z}{1 + c \cdot h_z}\right)}^2 \left(1 + c \cdot h_z\right)^2} \overset{!}{=} 0.03 \quad .
    \label{eq:hz-init-condition}
\end{equation}
An example of the fit procedure is shown in \cref{fig:magnetization-dependent-on-hz}
for a $4\times4$ cluster.
In this way, we determine a suitable value of the auxiliary staggered field $h_z$,
further details are given in App.\ \ref{sec:finitemag}. 

For the time evolution of the magnetization, the ground state  is initialized with the calculated auxiliary $h_z$ field. For $t>0$, $h_z$ is turned off and 
the switching field $h_y$ is 
applied to the system. Then, the Hamiltonian  reads
\begin{equation}
\label{eq:HamiltonianTimeEvolution}
\mathcal{H}_\mathrm{t} = -J \sum_{\substack{<i,j>}} \left[ \frac{\chi}{2} 
\left(S_i^+ S_j^+ + S_i^- S_j^- \right) + S_i^z S_j^z \right] - h_y \sum_{i} S_i^y .
\end{equation}
Note that the field in $y$ direction is uniform before and after the sublattice rotation
because the rotation was done around the $y$ direction so that this spin component is not
changed.

Since the ground state lies within the subspace $H_{(0, 0)}$ and the Hamiltonian $\mathcal{H}_t$ is also translationally invariant, the time evolution itself is restricted to this subspace. The time evolution is calculated using QuSpin \cite{weinberg17, weinberg19} by solving the time dependent Schr\"odinger equation using a Runge Kutta method, called dop853, of order 8.

\begin{figure}
    \centering
    \includegraphics[width=\columnwidth]{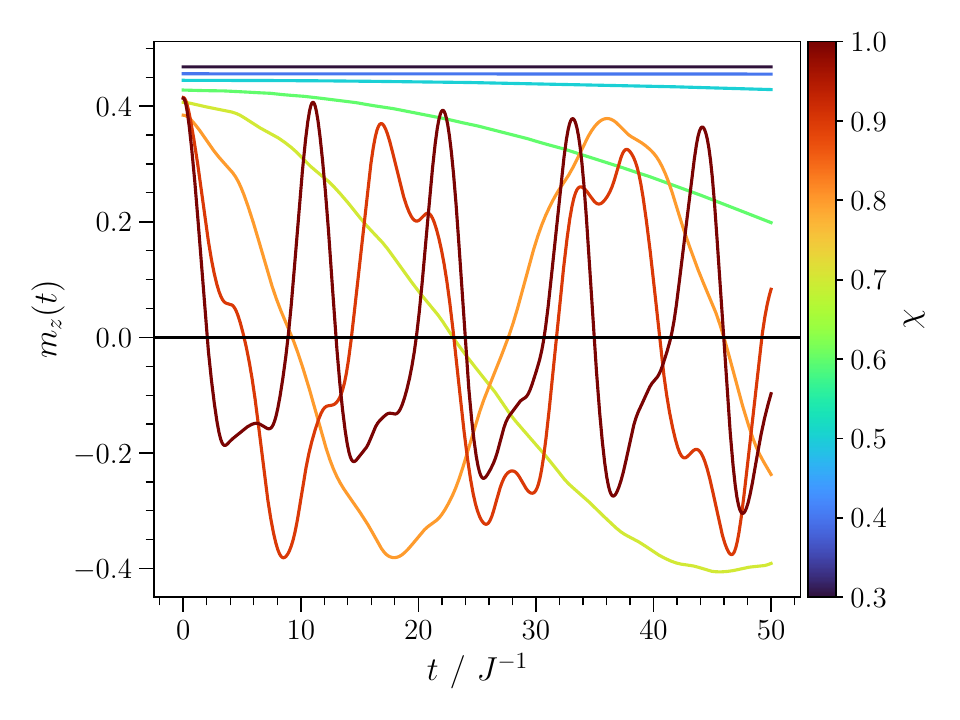}
    \caption{Spurious dynamics of the magnetization for a $4 \times 4$ cluster 
    after the auxiliary $h_z$ field has been turned off at $t > 0$. We point out that  this spurious evolution
    becomes progressively slower with increasing system size because the required auxiliary fields
    become smaller.}
    \label{fig:4x4-state-decay}
\end{figure}

Since the $h_z$ field is turned off for $t>0$ the initial state is no longer an eigenstate,
but will display some time dependence which we consider spurious. It cannot be avoided
unfortunately since we need initial states with finite sublattice magnetization. This
spurious time dependence is rather slow, at least for small enough values of 
$\chi$ where the quantum tunneling is weak, see Fig.\ \ref{fig:4x4-state-decay}. 
This effect is not very detrimental because the actual switching will be significantly faster.
We will focus on intermediate values of anisotropy $\chi\in (0.3, 0.8)$ where the 
quantum tunneling, sketched in Fig.\ \ref{fig:quantum-tunneling}, 
is indeed weak for the accessible cluster sizes.

\begin{figure}
    \centering
    \includegraphics[width=\columnwidth]{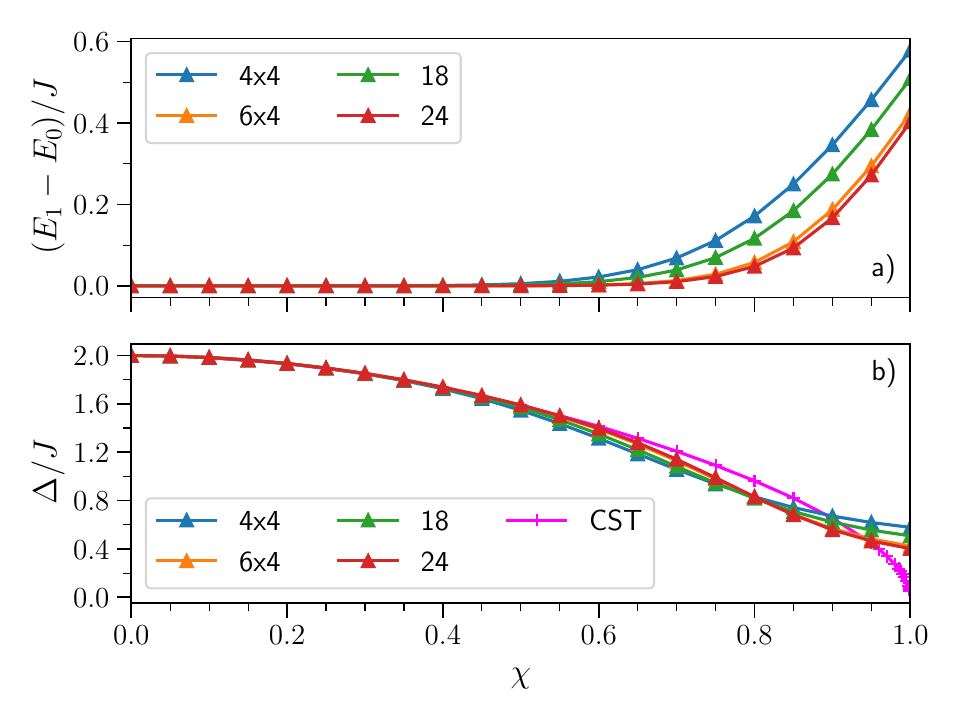}
    \caption{Results of ED calculations: panel a) Energy difference between the ground state
    $E_0$  and the first excited $E_1$ state in the subspace of total spin $S = 0$. Panel b) 
    Minimal energy required for one spin flip, i.e., for $\Delta S=1$,  compared to CST data \cite{caci24}. In both cases, no external $h_z$ field is applied. The clusters are 
    given explicitly in App.\ \ref{sec:lattice}.}
    \label{fig:energy-gaps}
\end{figure}

To quantify the degree of quantum tunneling we show the energy gap between the two
lowest eigenstates with the same total spin $S=0$ in the upper panel of Fig.\ 
\ref{fig:energy-gaps}. The matrix element of quantum tunneling is proportional
to the energy splitting of the two lowest eigenvalues. It becomes negligible for
small values of $\chi$ and decreases for larger clusters. Panel b)
displays the conventional spin gap for comparison which is accompanied by a change
of spin from $S=0$ to $S=1$.

Through the additional $h_y$ field the magnetization is switched from $m_z > 0$ to $m_z < 0$.  Our condition for the threshold of the switching field is set to be the $h_y$ value at which the first minimum of the magnetization curve touches $m_z=0$. This is done by using a Brent bisection method. In each iteration of $h_y$ the time evolution of the magnetization is calculated
for discrete time points, then interpolated by a cubic spline
and the first minimum of that function is computed.

\section{Results} \label{sec:results}

We simulate various cluster of rectangular and of rhombic shape, see App.\ \ref{sec:lattice}. During the initialization phase, we use the same $h_z$ fields  for both ED and SBMFT which depends on the anisotropy parameter $\chi$ to compare and to conduct the analyses of the system's dynamics induced by a switching field $h_y$.

The ground state of a finite quantum antiferromagnet without any external field applied,
as determined by ED, is a superposition of all the states in the subspace of total magnetization $M = 0$. 
This can be viewed as the effect of quantum tunneling. Thus, ED is unable to demonstrate switching in a 
direct manner because of the lack of spontaneous magnetization.
We incorporate an explicit symmetry-breaking term, which is achieved by introducing a Zeeman term, with a reasonably chosen small auxiliary $h_z$ field. Consequently, the cluster is initialized with non-degenerate low-energy states displaying finite sublattice magnetization.

\subsection{The dynamics of the magnetization}

\begin{figure}
    \centering
    \includegraphics[width=\columnwidth]{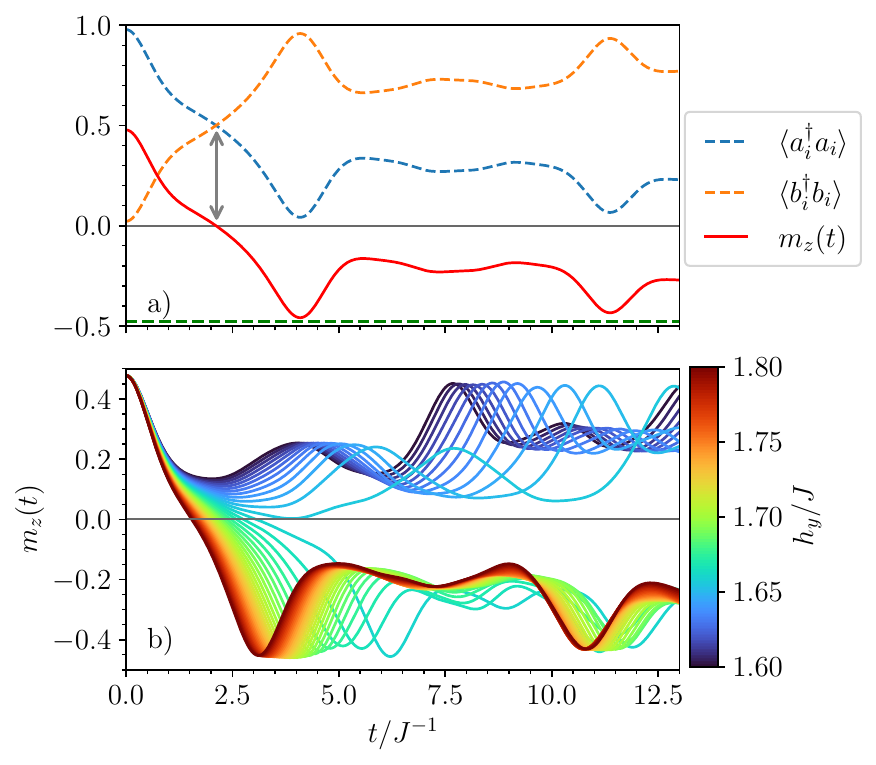}
    \caption{Panel (a) shows the dynamics of the occupation of the Schwinger bosons 
    $\lara{\nbi{a}}$ and $\lara{\nbi{b}}$ together with the resulting magnetization $m$ 
    from Eq.\ \eqref{eq:magdynam} for $h_y=1.7\,J$. The dashed green line shows the 
    negative value of the initial magnetization $m_{z0}=m_z(t=0)$. 
    Panel (b) represents the dynamics of the magnetization for external switching fields 
    in the interval $h_y\in [1.6 \,J,1.8\,J]$. Both graphs are the results of
    time-dependent SBMFT at $\chi=0.5$.}
    \label{fig:onlysbmft}
\end{figure}

To set the stage, we present the results of SBMFT for a relatively large system of 
$500\times500$ sites. Figure \ref{fig:onlysbmft}(a) 
shows the dynamics of the occupation numbers of $a$ 
and $b$ bosons together with the resulting sublattice magnetization according to
Eq.\ \eqref{eq:magdynam}. As mentioned in the Sect. \ref{subsec:sbmft}, there are  initially 
more $a$ bosons than $b$ bosons in the sublattice. To ensure proper convergence of self-consistency equations, we apply a small 
auxiliary field $h_z=1.329 \,J\,N^{-1}$\cite{bolsm23} initially. 
Then, the external driving field 
$h_y=1.7 \,J$ manipulates the bosonic occupation number in the sublattice such that more $b$ bosons replace $a$ bosons while keeping the total number constant. The transition of magnetization
takes place at the time point (vertical double-headed arrow in \cref{fig:onlysbmft}(a)) 
when these two bosonic flavors attain equal occupation. At later times, the magnetization becomes negative, i.e., the order switches from the $\uparrow$ state to the $\downarrow$ state, 
but with slowly decreasing oscillations after the switching.
We attribute these evanescent oscillations to dephasing effects \cite{khudoy25scipost}. 
This is one of the key findings of the time-dependent SBMFT, particularly in the context of 
studying switching, as SBMFT is capable of capturing the quantum effect of dephasing. 
This effect does not occur in classical macrospin calculations. 

Another important result is that magnetization never recovers its initial value in absolute terms. Its absolute value 
(the dashed green line in \cref{fig:onlysbmft}(a)) is not reached after reorientation because the control field increases the total energy of the system so that the magnetization
is pushed away from its initial values belong to its almost degenerate ground 
states. Note that the current model does not comprise any relaxation mechanisms \cite{khudoy25scipost}.

Figure \ref{fig:onlysbmft}(b) illustrates the dynamics of the sublattice magnetization in 
response to an external magnetic field from the interval of $h_y\in [1.6 \,J,1.8\,J]$ at 
$\chi=0.5$. It is easy to see that for small $h_y$ fields no switching is obtained, whereas 
high switching fields result in a sign change in the sublattice magnetization. It has been established that a threshold value of the field, denoted as $h_y^\mathrm{thr}$, 
is required to induce the desired transition \cite{bolsm23,khudoy24,khudoy25,cheng15}.We emphasize that it is an interesting property of this switching process 
that the sublattice magnetization stays switched even though the switching fields $h_y> h^\mathrm{thr}_y=1.656 \,J$ is kept at all times $t>0$.
One could have expected that the sublattice magnetization  starts to oscillate and keeps switching back and forth as would be the case for two macrospins \cite{nowak23,gomon10,khudoy24}. The reason why this does not occur is dephasing which is important in large quantum 
systems because of the destructive interference of a macroscopic number of modes
at all wave vectors $\mathbf{k}$. Dephasing is less pronounced in small quantum systems since there is not a large number of modes which can be superposed to reach destructive interference. 
Hence, one can expect that smaller quantum systems display more oscillations which do not fade away completely. This is particularly likely in the mean-field approach in which 
there is only one mode per momentum while in the full calculation the interplay of a huge number of multiple modes per momentum leads to enhanced interference.

\begin{figure} 
    \centering
    \includegraphics[width=\columnwidth]{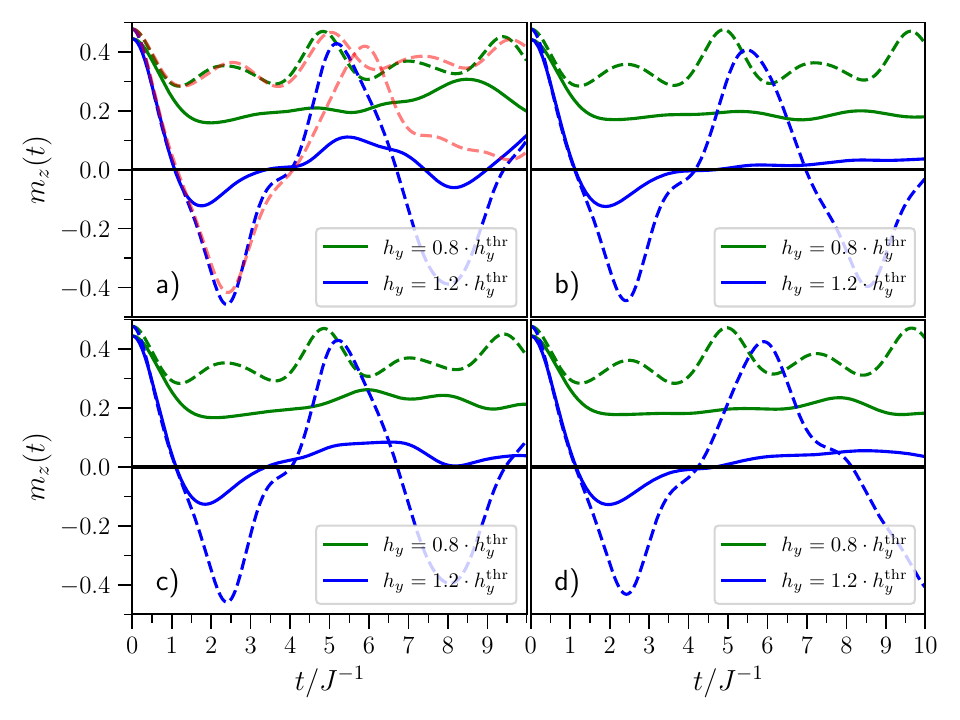}
    \caption{Comparison of the magnetization dynamics of  the results of ED (solid) and SBMFT (dashed) methods at $\chi = 0.5$. 
    a) 4x4, b) 6x4, c) 18, d) 24. The $h_y$ fields were chosen 
    with respect to the threshold fields $h^\mathrm{thr}_y$ of the respective method. 
    The red curve corresponds to the result of a $500 \times 500$ SBMFT cluster for comparison.}
    \label{fig:dynamics-chi-0.5}
\end{figure}

\begin{figure}
    \centering
    \includegraphics[width=\columnwidth]{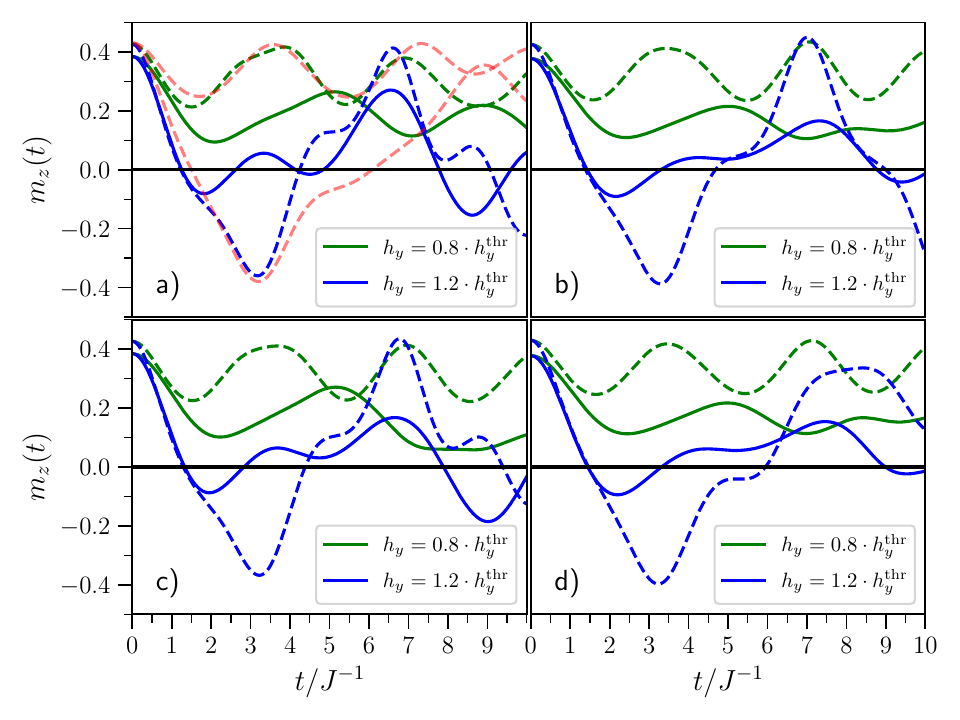}
    \caption{Comparison of the magnetization dynamics of the results of ED (solid) and 
    SBMFT (dashed) methods at $\chi = 0.8$. a) 4x4, b) 6x4, c) 18, d) 24. The $h_y$ fields were chosen 
    with respect to the threshold fields $h^\mathrm{thr}_y$ of the respective method. 
    The red curve corresponds to the result of a $500 \times 500$ SBMFT cluster for comparison.}
    \label{fig:dynamics-chi-0.8}
\end{figure}

 In Figs.\ \ref{fig:dynamics-chi-0.5} and \ref{fig:dynamics-chi-0.8} the time evolution of the sublattice magnetization obtained from ED and SBMFT is depicted for the different clusters for $\chi=0.5$ and $\chi=0.8$, respectively. In all cases, the $h_y$ values are chosen with respect to the threshold field of the corresponding approach, i.e., ED or SBMFT.
The green curves refer to fields $20$ \% below and the blue curves to fields $20$ \% above the switching threshold values of the respective method. 
In interpreting the shown data, one should keep in mind that there is some
spurious dynamics even without switching field due to the turned off auxiliary field, 
see \cref{fig:4x4-state-decay}. Obviously, this spurious dynamics is negligible for 
$\chi=0.5$ for the times displayed in Figs.\ \ref{fig:dynamics-chi-0.5} and \ref{fig:dynamics-chi-0.8}. But for $\chi=0.8$, this effect is not completely negligible,
at least for the $4\times 4$ cluster.

A comparison of the time evolution of magnetization resulting from the two approaches reveals similarities and differences. 
Figures \ref{fig:dynamics-chi-0.5} and \ref{fig:dynamics-chi-0.8} display the dynamics of the magnetization for the results of the ED (solid lines) and the SBMFT (dashed lines). The initial dynamics up to $t\approx 2/J$ is in good agreement and we will analyze it in more detail below. For longer times, however, deviations prevail. The ED results are considerably flatter, i.e.,
they do not show strong oscillations after the switching field has been turned on, while
the SBMFT results oscillate significantly stronger on the displayed time scales.
We attribute this difference to the difference in  interference. 
The ED calculations allow for the superposition for tremendously more eigen modes since the full Hilbert space increases exponentially  with system size. In contrast, the mean-field results only allow for interference of modes characterized by their wave number. Thus, this number of modes only increases proportional to the system size. Note that the increase of the system size in panels a)
does not have a large impact up to moderate times $\approx 5/J$. But for longer times
the increased destructive interference is discernible.

\subsection{The magnetic field threshold for switching} \label{sec:threshfield}

\begin{figure}
    \centering
    \includegraphics[width=\columnwidth]{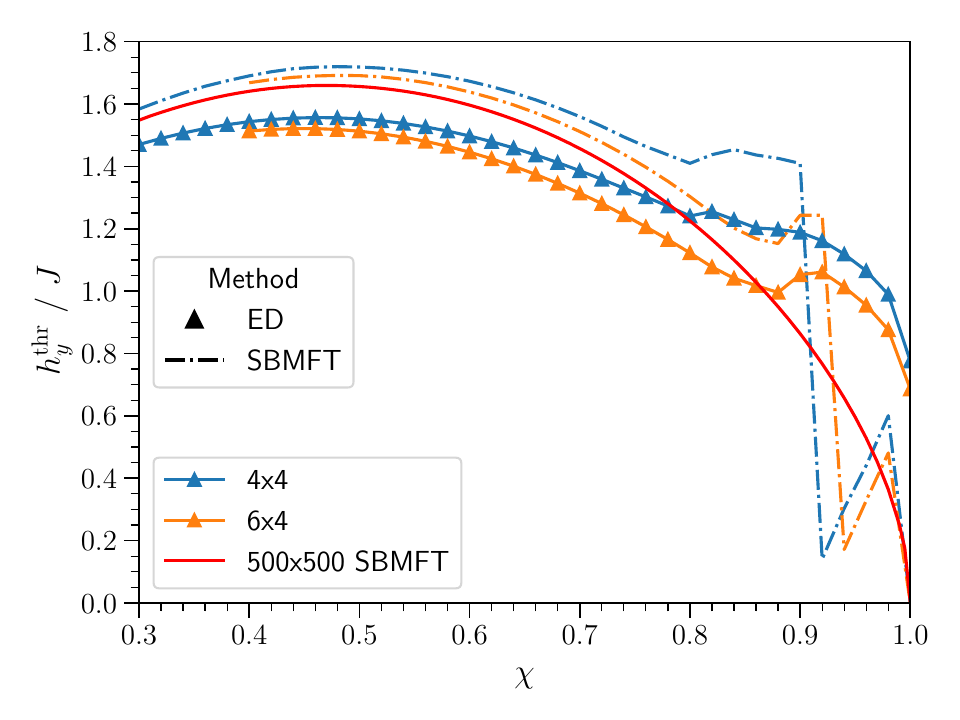}
    \caption{Comparison of the threshold fields depending on the anisotropy factor 
    $\chi$ for the $4 \times 4$ and the $6 \times 4$ spins clusters. The data for $\chi$ larger
    than $0.8$ ($0.88$) is strongly affected by finite-size effects for the $4\times4$ 
    ($6\times4$) cluster.}
    \label{fig:hy-thresholds-rectangular}
\end{figure}

Here we investigate the required minimum external field, to which  we  refer to as
``threshold'' field $h_y^\mathrm{thr}$, to achieve the desired switching of the sublattice magnetization. For a proper comparison, the estimated auxiliary $h_z$ fields from the exact diagonalization are also used for the Schwinger boson mean-field theory to determine the switching thresholds. In \cref{fig:hy-thresholds-rectangular} the obtained switching thresholds for the rectangular clusters are shown. The data from the SBMFT are drawn as solid lines. 
A comparison to a very large system of $500\times500$ sites is depicted as a red solid line.

Within the range of $\chi \in [0.4, 0.8]$ where both methods are well applicable, they yield the 
same shape with a nearly constant deviation. At $\chi = 0.8$ the threshold field for the $4\times4$ lattice suddenly starts to increase, which was observed for both methods. We attribute this to strong finite-size effects so that this region is not physically relevant.
If the system size is increased to $6\times 4$ spins, the sudden increase of the threshold 
field is shifted to a higher value of $\chi \approx 0.88$. This tendency corroborates the view that these artifacts vanish for an infinitely large system.
We expect that the shape of the threshold curve follows the shape of the red solid curve
obtained from SBMFT for a $500\times500$ system. In the range $\chi\in[0.4,0.8]$ the results of the exact diagonalization and the Schwinger boson mean-field theory are in good agreement.
The deviation is in the range of about $12.5\%$. This finding strongly supports the 
applicability of SBMFT to obtain a good understanding of the temporal evolution
on a semi-quantitative level.

\begin{figure}
    \centering
    \includegraphics[width=\columnwidth]{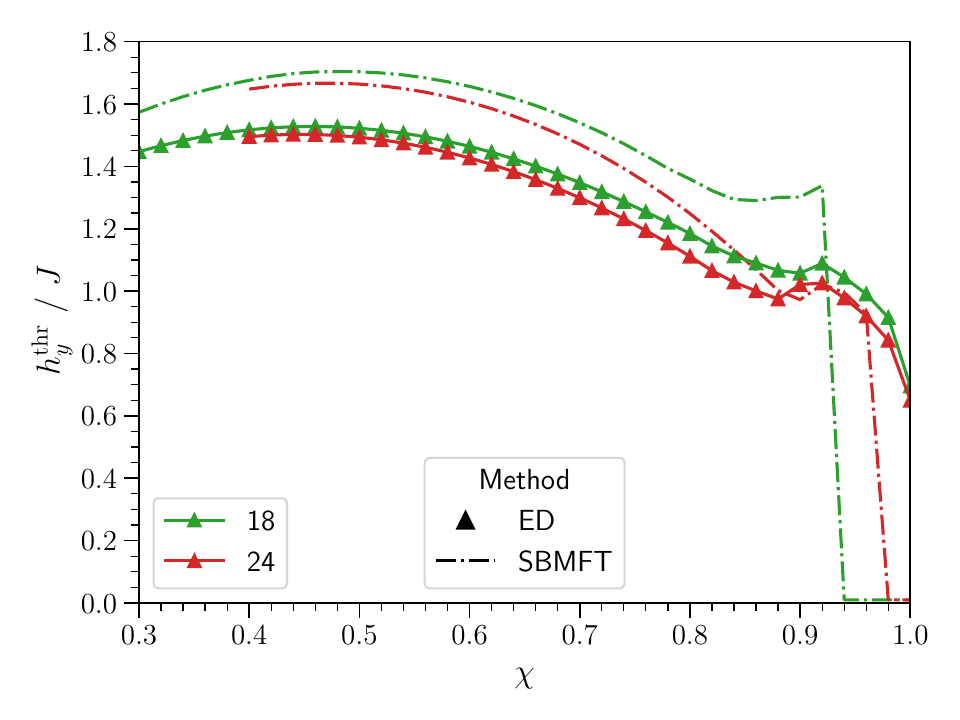}
    \caption{Comparison of the threshold fields depending on the anisotropy factor 
    $\chi$ for the $18$ and the $24$ rhombohedral spin clusters. The data for $\chi$ larger
    than $0.8$ ($0.88$) is strongly affected by finite-size effects for the $18$ 
    ($24$) cluster.}
    \label{fig:hy-thresholds-diamond}
\end{figure}

Inspecting the rhombohedral clusters, see App.\ \ref{sec:lattice}
for their shape, in \cref{fig:hy-thresholds-diamond}  the same key points can be noticed. 
 The shape of the threshold curve between $\chi \in [0.4, 0.88]$, as illustrated in \cref{fig:hy-thresholds-diamond}, is reproduced by the SBMFT with an  accuracy
 comparable to the one for the rectangular clusters. Also the strong finite-size effects set
 in at similar values of the anisotropy. These findings corroborate the above conclusion
 that SBMFT is a well-applicable semi-quantitative approach for switching processes.
 
 We add an explanation for the perhaps puzzling finite threshold field at $\chi=1$ where
 one naively expects that no activation energy needs to be overcome. This is indeed true
 for infinite systems \cite{bolsm23}. But for the finite clusters there is always a finite
 spin gap, see panel b) of \cref{fig:energy-gaps}, and hence always an activation energy
 to be overcome. Moreover, we point out that we use auxiliary fields to ensure
 a finite sublattice magnetization in the initial state. This puts the expectations
 for the isotropic antiferromagnetic clusters into perspective.

\subsection{Switching times from ED and SBMFT} 
\label{sec:threshtime}

\begin{figure}
    \centering
    \includegraphics[width=\columnwidth]{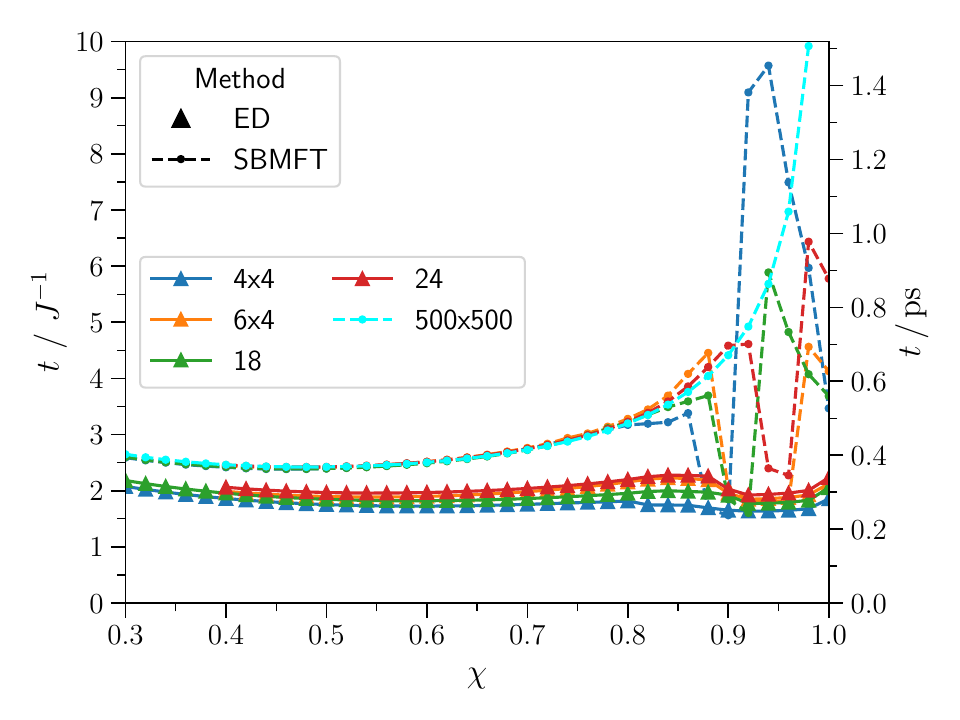}
    \caption{Switching times depending on the anisotropic factor $\chi$ 
    for the different clusters, where the switching $h_y$ are
    chosen 20 \% higher than the respective threshold fields $h^\text{thr}$ of the 
    method used.
    The $y$ axis on the right  indicates the switching times in picoseconds 
    assuming a coupling strength of $J = 10\, $ meV.}
    \label{fig:switching-times}
\end{figure}

In \cref{fig:switching-times} the switching times of the ED and SBMFT are shown.
Besides the minimal required fields $h_y^\text{thr}$ these times are relevant and characteristic 
quantities for the manipulation of the antiferromagnetic order parameters.
We determine the switching times form the first minimum of $m_z(t)$ once $m_z(t)$ has undergone
a sign change, see Figs.\ \ref{fig:dynamics-chi-0.5} and \ref{fig:dynamics-chi-0.8}.
In order not to be influenced by irrelevant fine-tuning we consider successful switching processes
by choosing fields $20$ \% above the threshold fields as determined for the respective method, ED or
SBMFT.

Passing from $\chi = 0.30$ to about $\chi = 0.60$ in the ED case, the switching times decrease
until they reach a minimum which differs slightly from cluster to cluster. 
If $\chi$ is increased further to $0.80$ the switching times start to rise again, until the
detrimental finite-size effects start to show up as discussed in the 
Fig.\ \ref{fig:hy-thresholds-rectangular}. 
In the large $N$ limit, i.e., in the thermodynamic limit, these effects are expected to vanish.
Thus, we expect that the dependence of the switching times on $\chi$ has a positive curvature 
with a minimum at around $\chi \approx 0.6$ for the infinite system.

For $\chi\to 1$, the SBMFT results for the large $500\times 500$ cluster indicate
 a divergence of the switching times. This is indeed realistic because in the isotropic limit
 the switching fields become smaller and smaller eventually vanishing for $\chi=1$. As a consequence
 the switching based on the Larmor precession about the switching field takes longer and longer so that the
 switching times diverge.
 This can be avoided by using staggered switching fields exploiting exchange-enhancement 
 \cite{khudoy25}. But this is beyond the scope of the present paper.

A comparison to the switching times of the SBMFT shows similar behavior. 
Passing from $\chi = 0.3$ up to roughly $\chi = 0.5$ the switching times decrease and 
start to increase as $\chi$ is becoming larger. At around $\chi = 0.8$ the switching times heavily oscillate as a consequence of the finite size clusters.
Keep in mind that the switching fields depend on $\chi$ as well.
This finite-size effect can also be observed for the other clusters at slightly larger 
$\chi$ values. The switching times are in reasonable agreement with the ED values with deviation of about 30\% for the larger clusters. In particular
the qualitative shape of their dependence is alike. Yet, the SBMFT yields 
higher switching times than the ED calculations. Inspecting Figs.\ \ref{fig:dynamics-chi-0.5} 
and \ref{fig:dynamics-chi-0.8}
the deviation can be understood. While the first zero of $m_z(t)$ occurs quite early and
matches very well between ED and SBMFT the subsequent minimum appears naturally later where
the results of both approaches differ. Since the ED curves are flatter overall their
minimum tends to occur earlier than the minimum in the SBMFT curve displaying larger
oscillations.

Since antiferromagnetic storage devices are expected to allow for ultrafast switching times
and hence for data manipulations in the range of THz, the estimated switching times are also converted to 
SI units. On the right $y$ axis of \cref{fig:switching-times}
the switching times are given in picoseconds assuming a coupling strength of $J = 10 \, \si{meV}$. 
In this case, they are in the range of $0.25-0.5$ ps for $\chi < 0.8$. Even if one assumes a coupling strength of only 
$J = 1\,\si{meV}$ the switching times would still be in the range 
of $2.5-5$ ps. Therefore, the ultra fast switching capabilities of antiferromagnets are
confirmed by both computational methods to be in the THz range.


\section{Conclusion} 
\label{sec:concl}

The presented theoretical results provide a comprehensive description of a switching process in 
quantum antiferromagnets by means of external magnetic fields. Using exact numerical calculations 
for small clusters, we have theoretically confirmed that the reorientation of the N\'eel vector 
is possible on an ultrafast time scale. Due to the finite size effects, the reliable
computation of the temporal evolution of the field-driven quantum state is 
limited to short time scales. 

Our comparison of the ED and SBMFT results shows that both approaches yield similar dynamic behavior in small systems
on short time scales. A minimum field strength is necessary to achieve switching.
The behavior of this threshold switching field strongly depends on the anisotropy of the system. 
The results for the threshold fields from ED and from SBMFT agree well with only about $12.5$\%
of deviation in the range of $\chi = [0.4-0.8]$. Moreover, the switching times are in picosecond regime 
in both approaches, but the differences between the methods are large of about
30\%. The switching time scale in the picosecond range
underlines the ultrafast magnetization dynamics of this manipulation of the long-range 
antiferromagnetic order parameter.  

The long time evolution, however, of the sublattice magnetization 
agrees only qualitatively between both approaches.  Due to the very large size of the Hilbert space, the 
ED calculations display stronger dephasing than the results from SBMFT, i.e., the ED curves are flatter.
The SBMFT results also show dephasing in contrast to macrospin calculations. But the number of modes
which contribute to a generic signal such as the sublattice magnetization grow only linear in system size.
On the one hand, this constitutes a certain caveat. On the other hand, 
it must be kept in mind that the ED approach is unable to reach the system sizes that would allow
reliable predictions for the thermodynamic limit.

In addition, the spin isotropic limit  $\chi\rightarrow 1$ cannot be analyzed by ED because the
finite-size effects become obstructively dominating. Yet the satisfactory agreement 
between SBMFT and ED for small systems and short times allows us to trust the physically reasonable 
SBMFT results in the isotropic limit \cite{bolsm23}. Furthermore, our previous studies demonstrated that 
the threshold fields can be significantly reduced based on exchange-enhancement \cite{khudoy25} and/or 
the application of time-dependent control fields \cite{khudoy24}. 

Although applying the theoretically found high switching fields is far from trivial and difficult 
to conduct experimentally in the laboratory, our results provide evidence that switching can be achieved
on picosecond time scales.

To conclude, the analyses in present article shows that the time-dependent Schwinger boson mean-field theory 
is a justified approach to treat the switching process in quantum antiferromagnets. The obtained results pave the
way for further investigations of switching processes and the description of non-classical behavior of the 
N\'eel vector in field-driven quantum antiferromagnets. This will help to leverage the potential of 
antiferromagnets for data storage and data processing. 

		
\begin{acknowledgments} 

We are grateful to Joachim Stolze for the helpful discussion.
This work has been financially supported by the Deutsche
Forschungsgemeinschaft (German Research Foundation)
in project UH 90/14-2.

\end{acknowledgments}

\appendix

\section{Finite sublattice magnetization ground state} \label{sec:finitemag}

Initializing a suitable ground state for the switching process is crucial in order to start with a 
finite sublattice magnetization state. This is why we  address this issue further. For small clusters,
such as the ones we are dealing with in this work, the ground state would always have neither 
a finite total magnetization nor a finite sublattice magnetization \cite{lieb62}
without an external magnetic field applied. This is why we have to apply a small auxiliary 
magnetic field $h_z$ in the $z$ direction in order to achieve a finite value of $m_z$. 
For a discrete set of $h_z$ values, the sublattice magnetization
\begin{equation}
    m_z(h_z) = \bra{\Psi_0} S^z_0 \ket{\Psi_0}
\end{equation}
is calculated. The state $\ket{\Psi_0}$ denotes the ground state of the system at the corresponding staggered $h_z$ field. 
The sublattice magnetization $m_z$ is only calculated for the spin with index $0$, because the lattice is translationally 
invariant after the sublattice rotation. The calculation of the expectation value for each spin 
in the cluster would become too time consuming for $N = 24$. 
For clarification we highlight, that after the rotation of one sublattice, all spins are aligned by the magnetic field 
in the same direction.  For the original unrotated system this corresponds
to a staggered, antiferromagnetic alignment of the spins. For each $\chi$, a few $m_z(h_z)$ values are calculated
as exemplified in Fig.\ \ref{fig:magnetization-dependent-on-hz}. Each blue dot corresponds to one ground state calculation 
performed in the $H_{(k_x, k_y)}$ subspace with $(k_x = 0, k_y = 0)$ which is done using QuSpin. 

In order to benefit from the translational invariance, we define equivalence classes.
For a given unitary symmetry transformation $Q$ with periodicity $M$, such that 
\begin{equation}
    Q^M = \mathbb{1} \quad ,
\end{equation}
the Hilbert space can be decomposed into equivalence classes, called cycles $C_i$. Each cycle
\begin{equation}
    C_i = \left\{ \ket{\phi_i}, Q \ket{\phi_i}, Q^2 \ket{\phi_i}, ..., Q^{L_i - 1} \ket{\phi_i} \right\}
\end{equation}
has length $L_i$ and one representative state $\ket{\phi_i}$
from which all other states in this cycle can be reconstructed by means of $Q$. 
For each quantum number $m \in [0, 1, ..., M-1]$ and representative $\ket{\phi_i}$, a new state
\begin{equation}
    \ket{m, \phi_i} = \frac{1}{L_i} \sum_{n = 0}^{L_i - 1} e^{-i (\frac{2 \pi m}{M}) n} \hat{Q}^n \ket{\phi_i}
\end{equation}
is defined. The corresponding eigenvalue equation reads
\begin{equation}
    Q \ket{m, \phi_i} = e^{i \frac{2 \pi m}{M}} \ket{m, \phi_i} \quad .
\end{equation}

Then, the subspace can be constructed by computing $\bra{m, \phi_i} \hat{H} \ket{m, \phi_j}$ for 
each quantum number $m$, using all different cycles $C_i$.
Once the subspace is formed, 
the ground state is searched by using the so called "eigsh" sparse eigensolver from the SciPy linear algebra library. 
Once the sublattice magnetization values are calculated for different $h_z$ fields, we use equation \eqref{eq:FitFunction} 
to fit the magnetization. Then the normalized derivative according to \eqref{eq:hz-init-condition} is 
calculated and the initial $h_z$ field is determined. 

\begin{figure} 
    \centering
    \includegraphics[width=\columnwidth]{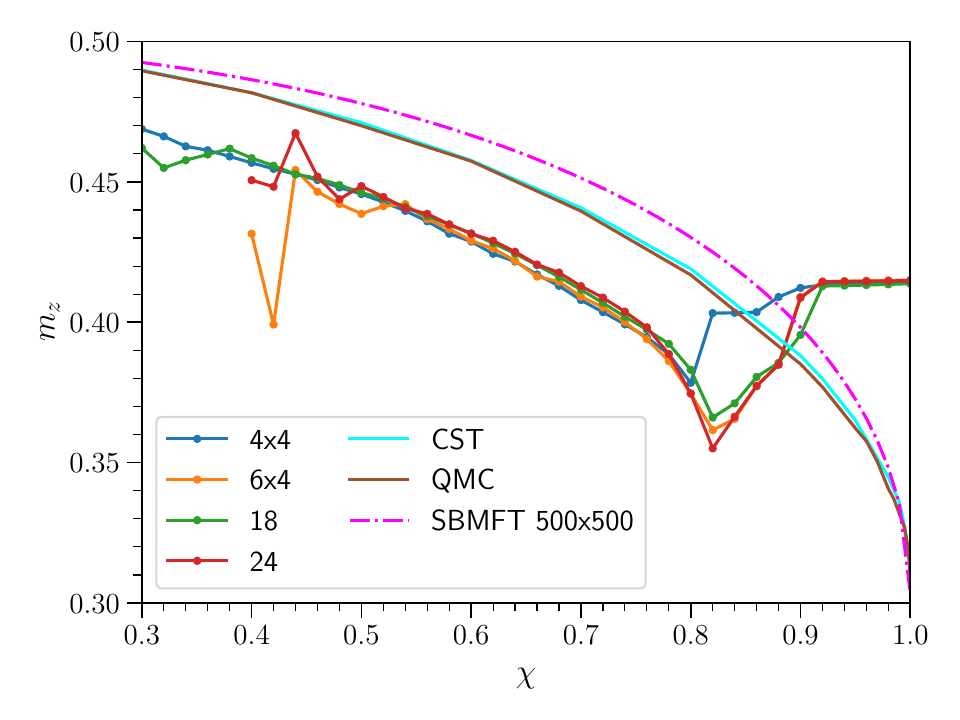}
    \caption{Comparison of the ground state sublattice magnetization between our finite clusters and the continuous 
    similarity transformation \cite{caci24} method and the quantum Monte Carlo \cite{sandv91, sandv99, sandv02}  
    approach in the thermodynamic limit without auxiliary $h_z$ field.}
    \label{fig:finite-magnetization-comparison}
\end{figure}

To determine a suitable value for the auxiliary field,
we require that the normalized derivative takes a value of $0.03$, 
yielding an auxiliary $h_z$ field value that is close to the point 
where the sublattice magnetization assumes its thermodynamics value.
Still, this condition is chosen a bit arbitrarily, but set in a way such that the 
initial sublattice magnetization agrees reasonably with results from other methods shown in \cref{fig:finite-magnetization-comparison}. 
This figure displays
a comparison between our estimated ground state sublattice magnetization to 
the results of a continuous similarity transformation (CST) \cite{caci24} and of a quantum Monte Carlo (QMC) \cite{sandv91, sandv99, sandv02} 
approach which provide results extrapolated to the thermodynamic limit. 
Our sublattice magnetization values underestimate the values of CST and QMC in the relevant low $\chi$ region.
Note, however, that the SBMFT result for a large system slightly overestimates  $m_z$ for all $\chi$ values. 
Summarizing, the curvature of the graph is reproduced well up to a value of approximately $\chi = 0.8$ 
where severe mismatch sets in due to dominant finite-size effects. 

\begin{figure}
    \centering
    \includegraphics[width=\columnwidth]{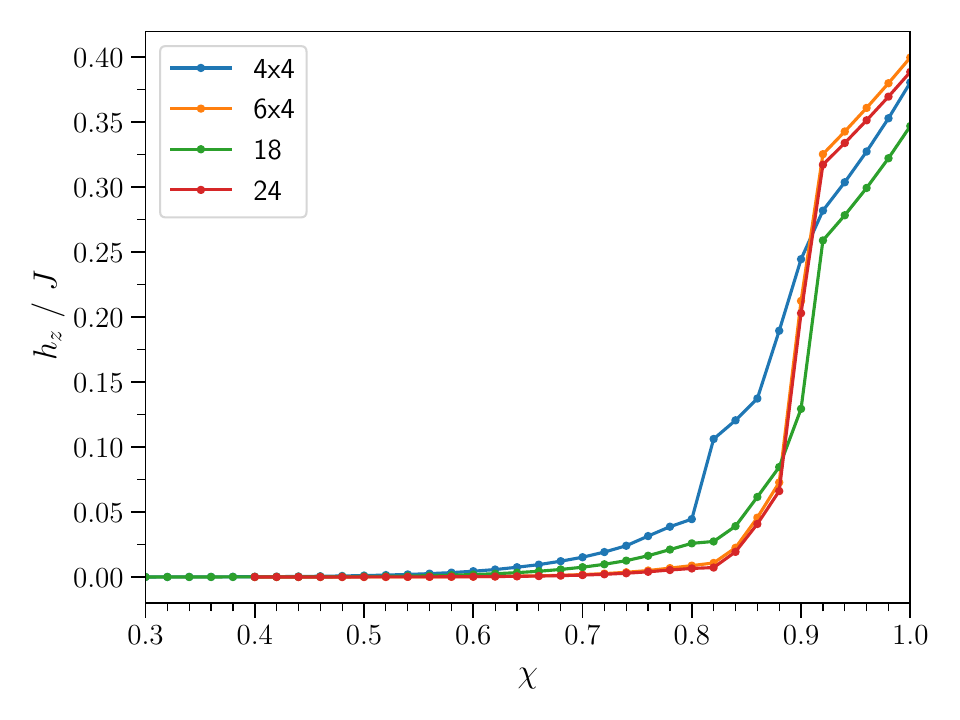}
    \caption{Estimated auxiliary $h_z$ fields as functions of the anisotropic factor $\chi$ for various clusters.}
    \label{fig:hz-initial}
\end{figure}

The resulting auxiliary magnetic fields are shown in \cref{fig:hz-initial}. For small values of $\chi$, 
only a very tiny field is required in order to induce a finite sublattice magnetization ground state. 
This is because the quantum tunneling from one ordered state to the other is strongly suppressed so that both are 
almost perfectly degenerate. If the system becomes more isotropic, larger and larger magnetic fields
are necessary to obtain the appropriate sublattice magnetization value.

\section{Lattice structures} \label{sec:lattice}

Here, we provide the used lattice structures for our calculations, as well as the corresponding \textbf{k} points in 
the reciprocal space. In every case periodic boundary conditions were used
indicated by the blue dots with primed numbers.

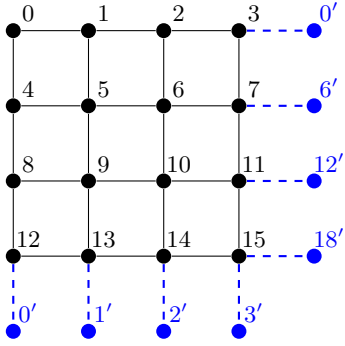
\begin{figure}
\centering
\begin{tikzpicture}[scale=1.0]
\def\Nx{4}  
\def\Ny{4}  

\foreach \i in {1,...,\Nx} {
  \foreach \j in {1,...,\Ny} {
    \pgfmathtruncatemacro{\index}{\i-1 + \Nx*(\j-1)}
    \node[circle,fill=black,inner sep=2pt,
          label={[shift={(0.2,-0.1)}]\index}] 
          (p\i\j) at (\i,-\j) {};
  }
}

\foreach \i in {1,...,\Nx} {
  \foreach \j in {1,...,\Ny} {

    \ifnum\i<\Nx
      \draw (p\i\j) -- (p\the\numexpr\i+1\relax\j);
    \fi

    \ifnum\j<\Ny
      \draw (p\i\j) -- (p\i\the\numexpr\j+1\relax);
    \fi
  }
}


\foreach \j in {1,...,\Ny} {
  \pgfmathtruncatemacro{\index}{(\j - 1)*6}
  \node[circle,fill=blue,inner sep=2pt, 
  label={[shift={(0.2,-0.1)}]\textcolor{blue}{$\index^{\prime}$}}] (ghostR\j) at (\Nx+1,-\j) {};
}

\foreach \i in {1,...,\Nx} {
  \pgfmathtruncatemacro{\index}{\i-1}
  \node[circle,fill=blue,inner sep=2pt, 
  label={[shift={(0.2,-0.1)}]\textcolor{blue}{$\index^{\prime}$}}] (ghostT\i) at (\i,-\Ny-1) {};
}


\foreach \j in {1,...,\Ny} {
  \draw[dashed, blue, thick] (p\Nx\j) -- (ghostR\j);
}

\foreach \i in {1,...,\Nx} {
  \draw[dashed, blue, thick] (p\i\Ny) -- (ghostT\i);
}

\end{tikzpicture}
\caption{4x4 spin lattice}
\[
k_x = -\pi + n \frac{\pi}{2}, \qquad n=0,1,2,3
\]
\[
k_y = -\pi + m \frac{\pi}{2}, \qquad m=0,1,2,3
\]
\label{fig:4x4-lattice}
\end{figure}



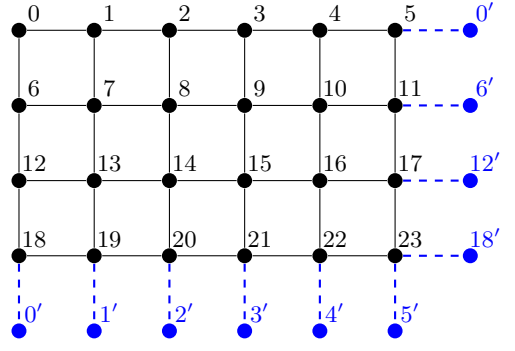
\begin{figure}
\centering
\begin{tikzpicture}[scale=1.0]
\def\Nx{6}  
\def\Ny{4}  

\foreach \i in {1,...,\Nx} {
  \foreach \j in {1,...,\Ny} {
    \pgfmathtruncatemacro{\index}{\i-1 + \Nx*(\j-1)}
    \node[circle,fill=black,inner sep=2pt,
          label={[shift={(0.2,-0.1)}]\index}] 
          (p\i\j) at (\i,-\j) {};
  }
}

\foreach \i in {1,...,\Nx} {
  \foreach \j in {1,...,\Ny} {

    \ifnum\i<\Nx
      \draw (p\i\j) -- (p\the\numexpr\i+1\relax\j);
    \fi

    \ifnum\j<\Ny
      \draw (p\i\j) -- (p\i\the\numexpr\j+1\relax);
    \fi
  }
}


\foreach \j in {1,...,\Ny} {
  \pgfmathtruncatemacro{\index}{(\j - 1)*6}
  \node[circle,fill=blue,inner sep=2pt, 
  label={[shift={(0.2,-0.1)}]\textcolor{blue}{$\index^{\prime}$}}] (ghostR\j) at (\Nx+1,-\j) {};
}

\foreach \i in {1,...,\Nx} {
  \pgfmathtruncatemacro{\index}{\i-1}
  \node[circle,fill=blue,inner sep=2pt, 
  label={[shift={(0.2,-0.1)}]\textcolor{blue}{$\index^{\prime}$}}] (ghostT\i) at (\i,-\Ny-1) {};
}


\foreach \j in {1,...,\Ny} {
  \draw[dashed, blue, thick] (p\Nx\j) -- (ghostR\j);
}

\foreach \i in {1,...,\Nx} {
  \draw[dashed, blue, thick] (p\i\Ny) -- (ghostT\i);
}

\end{tikzpicture}
\caption{6x4 spin lattice}
\[
k_x = -\pi + n \frac{\pi}{2}, \qquad n=0,1,2,3,4,5
\]
\[
k_y = -\pi + m \frac{\pi}{2}, \qquad m=0,1,2,3
\]
\label{fig:6x4-lattice}
\end{figure}



\begin{figure}
\centering
\begin{tikzpicture}[scale=1.0]

\node (p0) at (0,0) [circle,fill=black,inner sep=2pt, label={[shift={(0.2,-0.1)}]0}] {};
\node (p1) at (1,0) [circle,fill=black,inner sep=2pt, label={[shift={(0.2,-0.1)}]3}] {};
\node (p2) at (2,0) [circle,fill=black,inner sep=2pt, label={[shift={(0.2,-0.1)}]7}] {};
\node (p3) at (3,0) [circle,fill=black,inner sep=2pt, label={[shift={(0.2,-0.1)}]10}] {};
\node (p4) at (4,0) [circle,fill=black,inner sep=2pt, label={[shift={(0.2,-0.1)}]14}] {};
\node (p5) at (5,0) [circle,fill=black,inner sep=2pt, label={[shift={(0.2,-0.1)}]17}] {};

\node (p6) at (1,1) [circle,fill=black,inner sep=2pt, label={[shift={(0.2,-0.1)}]6}] {};
\node (p7) at (2,1) [circle,fill=black,inner sep=2pt, label={[shift={(0.2,-0.1)}]9}] {};
\node (p8) at (3,1) [circle,fill=black,inner sep=2pt, label={[shift={(0.2,-0.1)}]13}] {};
\node (p9) at (4,1) [circle,fill=black,inner sep=2pt, label={[shift={(0.2,-0.1)}]16}] {};

\node (p10) at (2,2) [circle,fill=black,inner sep=2pt, label={[shift={(0.2,-0.1)}]12}] {};
\node (p11) at (3,2) [circle,fill=black,inner sep=2pt, label={[shift={(0.2,-0.1)}]15}] {};

\node (p12) at (1,-1) [circle,fill=black,inner sep=2pt, label={[shift={(0.2,-0.1)}]1}] {};
\node (p13) at (2,-1) [circle,fill=black,inner sep=2pt, label={[shift={(0.2,-0.1)}]4}] {};
\node (p14) at (3,-1) [circle,fill=black,inner sep=2pt, label={[shift={(0.2,-0.1)}]8}] {};
\node (p15) at (4,-1) [circle,fill=black,inner sep=2pt, label={[shift={(0.2,-0.1)}]11}] {};

\node (p16) at (2,-2) [circle,fill=black,inner sep=2pt, label={[shift={(0.2,-0.1)}]2}] {};
\node (p17) at (3,-2) [circle,fill=black,inner sep=2pt, label={[shift={(0.2,-0.1)}]5}] {};

\foreach \i/\j in {
    0/1, 1/2, 2/3, 3/4, 4/5,
    6/7, 7/8, 8/9,
    10/11,
    12/13, 13/14, 14/15,
    16/17,
    6/1, 1/12,
    10/7, 7/2, 2/13, 13/16,
    11/8, 8/3, 3/14, 14/17,
    9/4, 4/15} {
  \draw (p\i) -- (p\j);
}

\node (p18) at (6,0) [circle,fill=blue,inner sep=2pt, 
    label={[shift={(0.2,-0.1)}]\textcolor{blue}{$0^{\prime}$}}] {};
\node (p19) at (5,1) [circle,fill=blue,inner sep=2pt, 
    label={[shift={(0.2,-0.1)}]\textcolor{blue}{$2^{\prime}$}}] {};
\node (p20) at (4,2) [circle,fill=blue,inner sep=2pt, 
    label={[shift={(0.2,-0.1)}]\textcolor{blue}{$1^{\prime}$}}] {};
\node (p21) at (3,3) [circle,fill=blue,inner sep=2pt, 
    label={[shift={(0.2,-0.1)}]\textcolor{blue}{$0^{\prime}$}}] {};
\node (p22) at (5,-1) [circle,fill=blue,inner sep=2pt, 
    label={[shift={(0.2,-0.1)}]\textcolor{blue}{$12^{\prime}$}}] {};
\node (p23) at (4,-2) [circle,fill=blue,inner sep=2pt, 
    label={[shift={(0.2,-0.1)}]\textcolor{blue}{$6^{\prime}$}}] {};
\node (p24) at (3,-3) [circle,fill=blue,inner sep=2pt, 
    label={[shift={(0.2,-0.1)}]\textcolor{blue}{$0^{\prime}$}}] {};

\foreach \i/\j in {
    5/18, 5/19, 9/19, 9/20, 11/20, 11/21,
    5/22, 15/22, 15/23, 17/23, 17/24} {
  \draw[dashed, blue, thick] (p\i) -- (p\j);
}

\end{tikzpicture}
\caption{18 spin lattice}
\[
k_x = \frac{ 2 \pi n}{6}, \qquad n=0,1,2,3,4,5
\]
\[
k_y = \frac{\pi}{3} \left( 2 m - n \right), \qquad m=0,1,2
\]
\label{fig:18-lattice}
\end{figure}
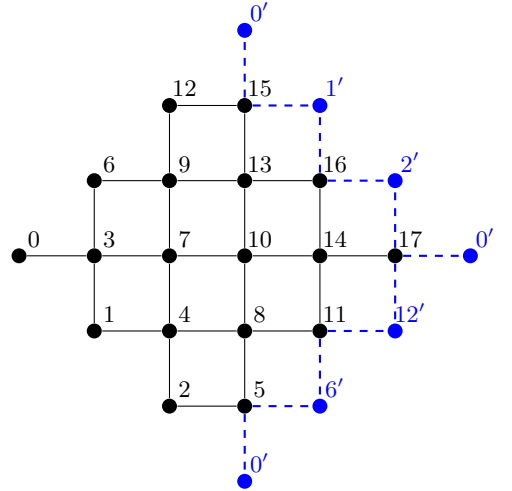



\begin{figure}[H]
\centering
\begin{tikzpicture}[scale=1.0]

\node (p0) at (0,0) [circle,fill=black,inner sep=2pt, label={[shift={(0.2,-0.1)}]0}] {};
\node (p1) at (1,0) [circle,fill=black,inner sep=2pt, label={[shift={(0.2,-0.1)}]3}] {};
\node (p2) at (2,0) [circle,fill=black,inner sep=2pt, label={[shift={(0.2,-0.1)}]7}] {};
\node (p3) at (3,0) [circle,fill=black,inner sep=2pt, label={[shift={(0.2,-0.1)}]10}] {};
\node (p4) at (4,0) [circle,fill=black,inner sep=2pt, label={[shift={(0.2,-0.1)}]14}] {};
\node (p5) at (5,0) [circle,fill=black,inner sep=2pt, label={[shift={(0.2,-0.1)}]17}] {};

\node (p6) at (1,1) [circle,fill=black,inner sep=2pt, label={[shift={(0.2,-0.1)}]6}] {};
\node (p7) at (2,1) [circle,fill=black,inner sep=2pt, label={[shift={(0.2,-0.1)}]9}] {};
\node (p8) at (3,1) [circle,fill=black,inner sep=2pt, label={[shift={(0.2,-0.1)}]13}] {}; 
\node (p9) at (4,1) [circle,fill=black,inner sep=2pt, label={[shift={(0.2,-0.1)}]16}] {};
\node (p10) at (5,1) [circle,fill=black,inner sep=2pt, label={[shift={(0.2,-0.1)}]20}] {};
\node (p11) at (6,1) [circle,fill=black,inner sep=2pt, label={[shift={(0.2,-0.1)}]23}] {};

\node (p12) at (2,2) [circle,fill=black,inner sep=2pt, label={[shift={(0.2,-0.1)}]12}] {};
\node (p13) at (3,2) [circle,fill=black,inner sep=2pt, label={[shift={(0.2,-0.1)}]15}] {};
\node (p14) at (4,2) [circle,fill=black,inner sep=2pt, label={[shift={(0.2,-0.1)}]19}] {};
\node (p15) at (5,2) [circle,fill=black,inner sep=2pt, label={[shift={(0.2,-0.1)}]22}] {};

\node (p16) at (3,3) [circle,fill=black,inner sep=2pt, label={[shift={(0.2,-0.1)}]18}] {};
\node (p17) at (4,3) [circle,fill=black,inner sep=2pt, label={[shift={(0.2,-0.1)}]21}] {};

\node (p18) at (1,-1) [circle,fill=black,inner sep=2pt, label={[shift={(0.2,-0.1)}]1}] {};
\node (p19) at (2,-1) [circle,fill=black,inner sep=2pt, label={[shift={(0.2,-0.1)}]4}] {};
\node (p20) at (3,-1) [circle,fill=black,inner sep=2pt, label={[shift={(0.2,-0.1)}]8}] {};
\node (p21) at (4,-1) [circle,fill=black,inner sep=2pt, label={[shift={(0.2,-0.1)}]11}] {};

\node (p22) at (2,-2) [circle,fill=black,inner sep=2pt, label={[shift={(0.2,-0.1)}]2}] {};
\node (p23) at (3,-2) [circle,fill=black,inner sep=2pt, label={[shift={(0.2,-0.1)}]5}] {};

\foreach \i/\j in {
    0/1, 1/2, 2/3, 3/4, 4/5,
    6/7, 7/8, 8/9, 9/10, 10/11,
    12/13, 13/14, 14/15,
    16/17,
    18/19, 19/20, 20/21,
    22/23,
    6/1, 1/18,
    12/7, 7/2, 2/19, 19/22,
    16/13, 13/8, 8/3, 3/20, 20/23,
    17/14, 14/9, 9/4, 4/21,
    15/10, 10/5} {
  \draw (p\i) -- (p\j);
}


\node (p24) at (6,0) [circle,fill=blue,inner sep=2pt, 
    label={[shift={(0.2,-0.1)}]\textcolor{blue}{$18^{\prime}$}}] {};
\node (p25) at (7,1) [circle,fill=blue,inner sep=2pt, 
    label={[shift={(0.2,-0.1)}]\textcolor{blue}{$0^{\prime}$}}] {};
\node (p26) at (6,2) [circle,fill=blue,inner sep=2pt, 
    label={[shift={(0.2,-0.1)}]\textcolor{blue}{$2^{\prime}$}}] {};
\node (p27) at (5,3) [circle,fill=blue,inner sep=2pt, 
    label={[shift={(0.2,-0.1)}]\textcolor{blue}{$1^{\prime}$}}] {};
\node (p28) at (4,4) [circle,fill=blue,inner sep=2pt, 
    label={[shift={(0.2,-0.1)}]\textcolor{blue}{$0^{\prime}$}}] {};
\node (p29) at (5,-1) [circle,fill=blue,inner sep=2pt, 
    label={[shift={(0.2,-0.1)}]\textcolor{blue}{$12^{\prime}$}}] {};
\node (p30) at (4,-2) [circle,fill=blue,inner sep=2pt, 
    label={[shift={(0.2,-0.1)}]\textcolor{blue}{$6^{\prime}$}}] {};
\node (p31) at (3,-3) [circle,fill=blue,inner sep=2pt, 
    label={[shift={(0.2,-0.1)}]\textcolor{blue}{$0^{\prime}$}}] {};

\foreach \i/\j in {
    23/31, 23/30, 21/30, 21/29, 5/29, 5/24,
    11/24, 11/25, 11/26, 15/26, 15/27, 17/27, 17/28} {
  \draw[dashed, blue, thick] (p\i) -- (p\j);
}

\end{tikzpicture}
\caption{24 spin lattice}
\[
k_x = \frac{\pi}{2} \left(\frac{n}{3} + \frac{m}{2} \right), \qquad n=0,1,2,3,4,5
\]
\[
k_y = \frac{\pi}{2} \left(\frac{m}{2} - \frac{n}{3} \right), \qquad m=0,1,2,3
\]
\label{fig:24-lattice}
\end{figure}
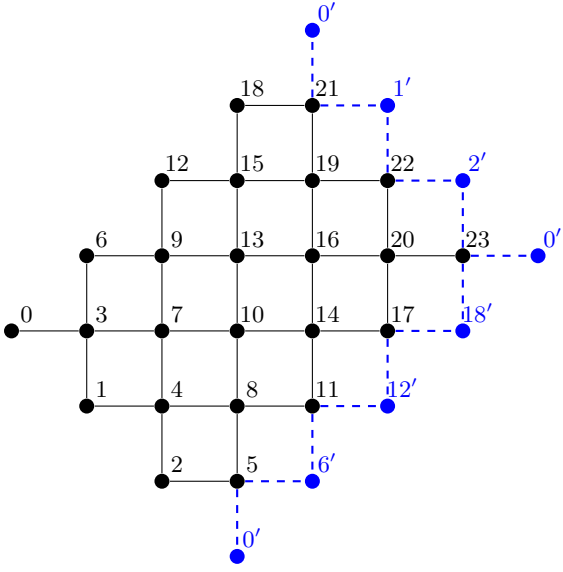  
 
\newpage

\bibliography{liter10.bib}

\end{document}